\documentclass[a4paper]{jpconf}
\usepackage{graphicx}
\usepackage{wrapfig}

\def\lsim{\mathrel{\lower4pt\hbox{$\sim$}} 
\hskip-9.5pt\raise1.6pt\hbox{$<$}\;} 
 
\def\gsim{\mathrel{\lower4pt\hbox{$\sim$}} 
\hskip-9.5pt\raise1.6pt\hbox{$>$}\;} 

\begin{document}
\title{The Inert Doublet Model : a new archetype of WIMP dark matter?}

\author{Michel H.G. Tytgat}

\address{Service de Physique Th\'eorique, CP225, Universit\'e Libre de Bruxelles, Bld du Triomphe 1050 Brussels, Belgium }

\ead{mtytgat@ulb.ac.be}

\begin{abstract}
The Inert Doublet Model (IDM) is a two doublet extension of the Higgs-Brout-Englert sector of the Standard Model with a $Z_2$ symmetry in order to prevent FCNC. If the $Z_2$ symmetry is not spontaneously broken, the lightest neutral extra scalar is a dark matter candidate. We briefly review the phenomenology of the model, emphasizing its relevance for the issue of Electroweak Symmetry Breaking (EWSB) and the prospects for  detection of dark matter.
\end{abstract}

\section{Introduction}

A host of dark matter candidates have been proposed over the years, some more natural than others. While the axion is an old contender, the most acclaimed candidate is the neutralino which, together with the so-called KK-photon and the heavy photon in Little Higgs models, belongs to the category of weakly interacting massive particles (WIMP). (See the review by Scopel at this conference \cite{Scopel:2007db}). Other models of dark matter are  often looked at with some disdain for, it is argued, they lack motivation. Well, supersymmetry is certainly well motivated, but the fact is that, due to constraints from colliders,  considerable fine tuning is required to obtain the right abundance of neutralinos.\footnote{By this I mean fine tuning in parameter space on top of the adjustment inherent to the WIMP paradigm. Even the simplest instance of dark matter, a stable heavy neutrino with weak interactions only, has an energy density that spans more than four decades (see figure 1 in \cite{Scopel:2007db}) above and below the observed value. Incidentally, why the latter is then of the same order of magnitude as the contribution of baryons is a mystery not addressed by the WIMP paradigm.} It is in our opinion fair to consider simpler, albeit phenomenologically interesting, toy models with dark matter. The Inert Doublet Model is such a model \cite{Deshpande:1977rw}. It is a simple extension of the Standard Model, with a Brout-Englert-Higgs sector that could impact electroweak symmetry breaking \cite{Barbieri:2006dq,Hambye:2007vf}, it is compatible with colliders constraints \cite{Babu:2007sm} and has a dark matter candidate that could be observable in the near future \cite{LopezHonorez:2006gr,Gustafsson:2007pc}. Furthermore, with minor generalization \cite{Ma:2006km}, it could establish a link between neutrino physics and dark matter.

\section{The IDM and Electroweak Symmetry Breaking}
The model we consider is a  two Higgs doublet extension of the SM, $H_1=(h^+ \, (h+iG_0)/\sqrt{2})^T$ and $H_2=(H^+ \, (H_0+iA_0)/\sqrt{2})^T$, 
together with a $Z_2$ symmetry such that 
all fields of the Standard Model and $H_1$  are even under $Z_2$ while
$
H_2\rightarrow - H_2. 
$
We assume that $Z_2$ is not spontaneously broken, {\em i.e.} that $H_2$ does 
not develop a {\it vev}. As there is no mixing between the doublets, $h$ plays the role of the usual Higgs particle. Because the extra doublet does not couple to quarks (and leptons in the simplest case) 
there are no FCNC.  
The most general renormalisable (CP conserving) potential of the model 
is 
\begin{eqnarray} 
\label{potential} 
V &=& \mu_1^2 \vert H_1\vert^2 + \mu_2^2 \vert H_2\vert^2  + \lambda_1 \vert H_1\vert^4 + 
 \lambda_2 \vert H_2\vert^4 \\ \nonumber 
&&\\
&&  + \lambda_3 \vert H_1\vert^2 \vert H_2 \vert^2 
 + \lambda_4 \vert H_1^\dagger H_2\vert^2 + {\lambda_5\over 2} \left[(H_1^\dagger H_2)^2 + h.c.\right] \nonumber
\end{eqnarray} 
with real quartic couplings.
The $SU(2) \times U(1)$ symmetry is broken by the vacuum expectation value of $H_1$, 
$\langle H_1\rangle = {v/\sqrt 2}
$ 
with $v = -\mu_1^2/\lambda_1 = 246$ GeV 
while, assuming $\mu_2^2 > 0$, 
$
\langle H_2\rangle = 0. 
$ 
\begin{wrapfigure}{r}{0.5\textwidth}
\vspace{-30pt}
  \begin{center}
    \includegraphics[width=0.45\textwidth]{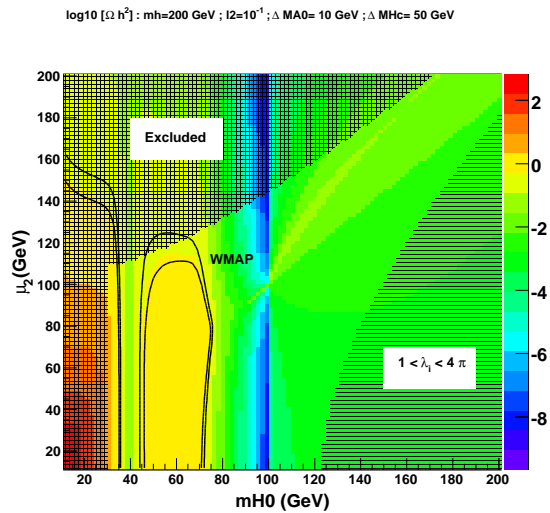}
  \end{center}
\vspace{-20pt}
  \caption{$\log_{10} \Omega h^2$ of $H_0$ for $M_h=200$ GeV, computed using micrOMEGAs2.0 \cite{Belanger:2006is}.}
\vspace{-10pt}
\end{wrapfigure}

We choose (this is arbitrary) $H_0$ to be the dark matter candidate. Two $H_0$ mass ranges are consistent with dark matter observations \cite{Barbieri:2006dq,LopezHonorez:2006gr,Cirelli:2005uq}, either $M_{H_0} \lsim 80$ GeV or $M_{H_0} \gsim 500$ GeV. The former is the most interesting from the point of view of phenomenology and observations. The relic abundance of $H_0$ is shown on Figure 1 below. 

The mass spectrum of the extra scalars depends on the quartic couplings $\lambda_3,\lambda_4$ and $\lambda_5$ and on the mass parameter $\mu_2$. Two interesting situations may arise. Firstly, if $\lambda_5$ is small, there is an approximate $U(1)_{PQ}$ symmetry and $M_{H_0} \approx M_{A_0}$ while, generically, $M_{H^\pm} \gg M_{H_0}, M_{A_0}$ (large isospin breaking). Less obvious is the custodial symmetry that arises if $M_{H^\pm} \approx M_{H_0}$ (resp. $M_{H^\pm} \approx M_{A_0}$) \cite{Gerard:2007kn}. Both limits are of phenomenological interest. 

First, large isospin breaking leads to large contributions to  $\rho=M_W^2/c^2_\theta M_Z^2$ or, equivalently, to the $T$ parameter \cite{Barbieri:2006dq}
$$
\Delta T  \approx {1\over 24 \pi^2 \alpha v^2} (M_{H^\pm} - M_{H_0})(M_{H^\pm} - M_{A_0})
$$
This feature has been exploited in  \cite{Barbieri:2006dq} to screen the $Z$ and $W^\pm$ gauge boson masses from an equally large (but opposite in sign) contribution from a heavy Higgs (achieving $M_h$ up to $500 GeV$). 

Second, we have exploited the custodial symmetry (vanishing contribution to $\Delta T$)  in \cite{Hambye:2007vf} together with the possibility that $M_{H^\pm} \approx M_{H_0} \ll M_{A_0}$ (resp. $M_{H^\pm} \approx M_{A_0} \gg M_{H_0}$) to give large radiative corrections to the effective potential. In particular, the large (negative) radiative effect of the top quark can be compensated by a large contribution from the extra scalars. In the extreme, although very suggestive, limit of vanishing $\mu_{1,2}$ parameters (the Coleman-Weinberg scenario), the electroweak symmetry breaking is literally induced by the extra scalar doublet {\em i.e.} by WIMP dark matter, a possibility that has been essentially overlooked in the litterature. Some working cases  are listed in Table 1. 

\begin{table}
\begin{minipage}[t]{0.58\linewidth}
\begin{tabular}{ |c|c|c||c|c|c|c||c|}\hline
  $\lambda_3$& $\lambda_4$ &$\lambda_5$ & $M_h$ & $M_{H_0}$ &  $M_{A_0}$ & $M_{H^\pm}$  \\
\hline\hline
  5.4 & -2.7 & -2.7 & 120 & 43 & 395 & 395 \\
\hline
  5.4 &  -2.6 & -2.6 & 120 & 72 & 390 & 390   \\
\hline
 7.6 & -4.1 & -4.1 & 180 & 12 & 495 & 495   \\
\hline
 7.6 & -3.8 & -3.8 & 180 & 64 & 470 & 470 \\
\hline
  -0.003 & 4.6 & -4.7 & 120 & 39 & 500 & 55   \\
\hline
  -0.07 & 5.5 & -5.53 & 150 & 54 & 535 & 63  \\
\hline
\end{tabular}
\end{minipage}
\hfill
\begin{minipage}[t]{0.38\linewidth}
\vskip -60pt
      \caption{Instances of mass spectrum from radiative symmetry breaking.  The quartic couplings are large, but  still in the perturbative regime. All these solutions are compatible with the observed abundance of dark matter.}
    \end{minipage}
\end{table}

\section{Direct and indirect detection}

Being a scalar particle, the cross-section for direct detection is 100 \% spin-independent (SI), a specific feature (one could say a limitation) of the model. In Figure 2 we show the prospect for direct detection  while  Figure 3 displays the integrated flux from the GC into   gamma rays \cite{LopezHonorez:2006gr}.  Gustafsson {\em et al} (see \cite{Gustafsson:2007pc} and the talk by Lars Bergstrom at this conference \cite{Bergstrom:2007rz}) have shown that annihilation into photon pairs, which occurs at one-loop, could give a significant gamma ray line. For $M_{H_0} \sim 70$ GeV say, this is in the window of opportunity of the forthcoming GLAST detector. As emphasized in  \cite{Gustafsson:2007pc} this enhancement ({\em i.e.} compared the neutralino signal) is a distinctive signal, due to the fact that the inert doublet does not couple to quarks. 

\begin{figure}
\begin{minipage}[b]{0.5\linewidth} 
\centering
\includegraphics[width=7cm]{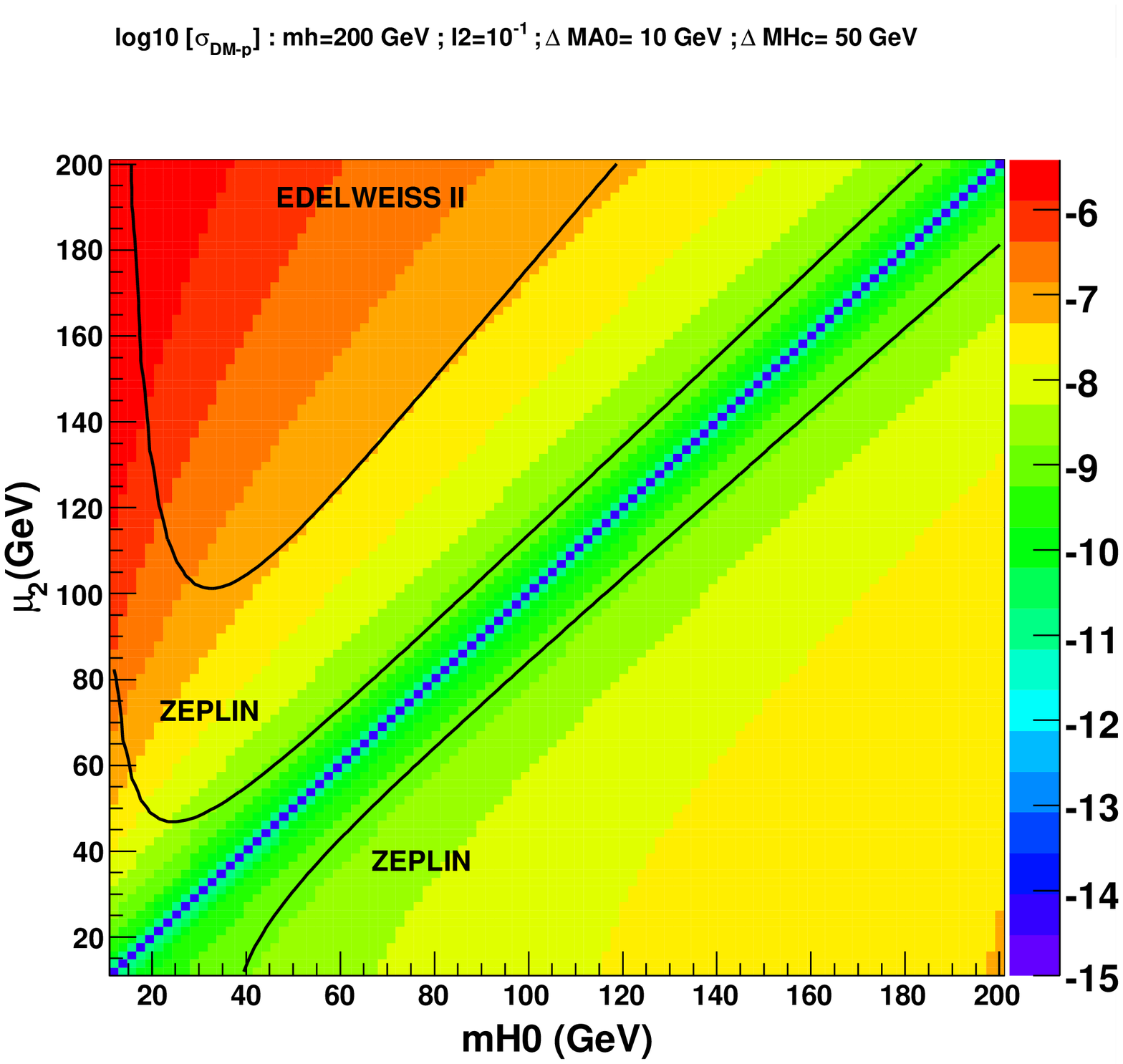}
\caption{$H_0$-nuclei cross-section with $M_h=200$}
\end{minipage}
\hspace{0.5cm} 
\begin{minipage}[b]{0.5\linewidth}
\centering
\includegraphics[width=7cm]{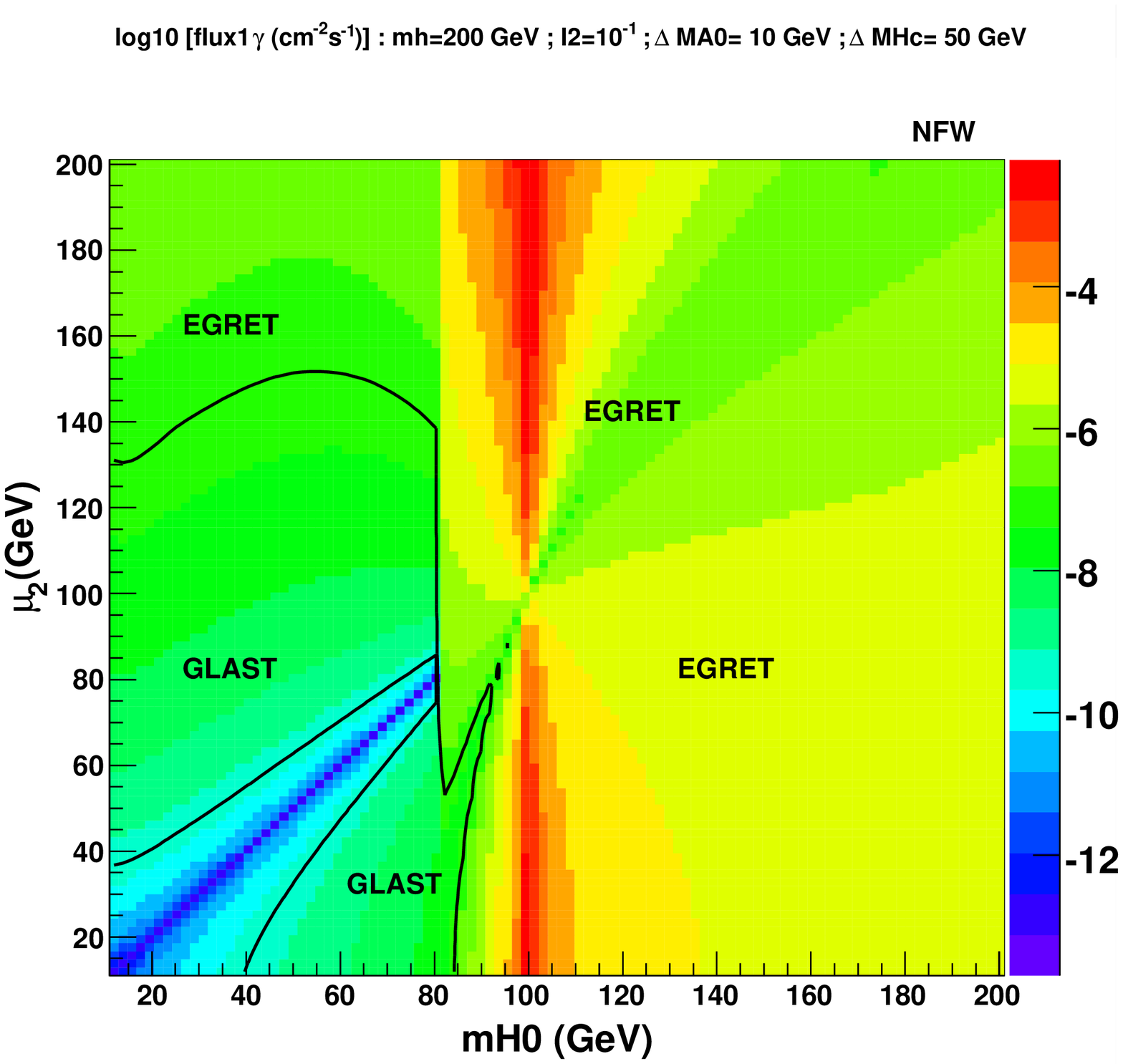}
\caption{Integrated gamma ray flux from $H_0$ annihilation with $M_h=200$ GeV.}
\end{minipage}
\end{figure}

\section*{Acknowledgements}
The author thanks the belgian FNRS and the IAP VI/11 for financial support and 
Laura Lopez Honorez, Emmanuel Nezri, Josep Oliver and Thomas Hambye for collaboration.

\end{document}